\documentstyle[12pt]{article}

\begin{document}

\begin{center}
{\bf SPONTANEOUS LORENTZ VIOLATION, GRAVITY \\
AND NAMBU-GOLDSTONE MODES 
\footnote{\tenrm
Presented at the 11th Marcel Grossmann Meeting,
Berlin, Germany, July 2006}
\\}
\end{center}

\begin{center}
{Robert Bluhm \\ }
{Physics Department \\ }
{Colby College \\ }
{Waterville, ME 04901 USA\\}
\end{center}

\begin{abstract}
\noindent
A brief summary is presented of recent work examining the fate
of the Nambu-Goldstone modes in gravitational theories with
spontaneous Lorentz violation.
\end{abstract}

\section{Introduction}	

The Goldstone theorem states that when a continuous global symmetry is
spontaneously broken, massless Nambu-Goldstone (NG) modes appear.
Alternatively, if the symmetry is local, 
then a Higgs mechanism can occur in which the gauge
fields acquire mass. 
In this work,
these processes are examined for the case where the symmmetry
is Lorentz symmetry
\cite{rbak05}.

\section{Spontaneous Lorentz Breaking}

Lorentz symmetry is spontaneously broken when a local tensor 
field acquires a vacuum expectation value (vev).
In curved spacetime, the Lorentz group acts locally at each spacetime point.  
In addition to being locally Lorentz invariant, a gravitational theory is also
invariant under diffeomorphisms.  
There are therefore two relevant symmetries, 
and it is important to consider them both.

One can show in general that when a vacuum expectation value
spontaneously breaks Lorentz symmetry,
then diffeomorphisms are also spontaneously broken.
The spontaneous breaking of these symmetries implies that NG
modes should appear (in the absence of a Higgs mechanism).
In general, there can be up to as many NG modes as there are broken symmetries. 
Since the maximal symmetry-breaking case would yield six broken Lorentz generators 
and four broken diffeomorphisms,
there can therefore be up to ten NG modes.
One natural gauge choice is to put all the NG modes into the vierbein,
as a simple counting argument shows is possible.
The vierbein has 16 components.  
With no spontaneous Lorentz violation, 
the six Lorentz and four diffeomorphism degrees
of freedom can be used to reduce the vierbein down to six independent degrees
of freedom.  
Thus,  in a theory with spontaneous Lorentz breaking,
up to ten NG modes can appear, and all of them can naturally
be incorporated as degrees of freedom in the vierbein.

The simplest example of a theory with spontaneous Lorentz breaking is 
a bumblebee model
\cite{ks,bb,note}. 
These are defined as theories in which a vector field $B_\mu$ acquires a vev,
$<B_\mu> \, = b_\mu$.     
The vev can be induced by a potential $V$ in the Lagrangian that has a minimum for 
nonzero values of the vector field.  
A simple example of a bumblebee model has the form
${\cal L} = {\cal L}_{\rm G} + {\cal L}_{\rm B} + {\cal L}_{\rm M}$,
where ${\cal L}_{\rm G}$
describes the pure-gravity sector,
${\cal L}_{\rm M}$ describes the matter sector,
and (choosing a Maxwell form for the kinetic term)
\begin{equation}
{\cal L}_{\rm B} = \sqrt{-g} \left( - \frac 1 4 B_{\mu\nu} B^{\mu\nu} 
- V(B_\mu) + B_\mu J^\mu \right) ,
\label{BBL}
\end{equation}
describes the bumblebee field.
Here,
$J^\mu$ is a matter current,
and the bumblebee field strength is
$B_{\mu\nu} = D_\mu B_\nu - D_\nu B_\mu$,
which in a Riemann spacetime (with no torsion) reduces to
$B_{\mu\nu} = \partial_\mu B_\nu - \partial_\nu B_\mu$.
(For simplicity, we are neglecting additional possible interactions between the
curvature tensor and $B_\mu$).

Among the possible choices for the potential are a sigma-model potential
$V = \lambda (B_\mu B^\nu \pm b^2)$, 
where $\lambda$ is a Lagrange-multiplier field,
and a squared potential
$V = \frac 1 2 \kappa (B_\mu B^\nu \pm b^2)^2$,
where $\kappa$ is a constant (of mass dimension zero).
In the former case,
only excitations that stay within the potential minimum
(the NG modes) are allowed by the constraint imposed by $\lambda$.
However,
in the latter case,
excitations out of the potential minimum are possible as well.
In either of these models, 
three Lorentz symmetries and one diffeomorphism are broken, 
and therefore up to four NG modes can appear.  
However, the diffeomorphism NG mode does not propagate
\cite{rbak05}.
It drops out of all of the kinetic terms and is purely an auxiliary field.  
In contrast, the Lorentz NG modes do propagate.  
They comprise a massless vector,
with two independent transverse degrees of freedom (or polarizations).
Indeed,
they are found to propagate just like a photon.

We find that the NG modes resulting from spontaneous local Lorentz violation 
can lead to an alternative explanation for the existence of massless photons
(besides that of U(1) gauge invariance).
Prevoius links between QED gauge fields, fermion composites, and the NG modes 
had been uncovered in flat spacetime (with global Lorentz symmetry). 
Here, we propose a theory with just a vector field (but no U(1) gauge symmetry)
giving rise to photons in the context of a gravitational theory
where local Lorentz symmetry is spontaneously broken.
Defining $B_\mu - b_\mu = A_\mu$, 
we find at lowest order that the Lorentz NG excitations propagate 
as transverse massless modes obeying an axial gauge condition,  
$b_\mu A^\mu = 0$. 
Hence, we conclude that
spontaneous local Lorentz violation may provide an alternative explanation for
massless photons.
In the bumblebee model,
the photon fields
couple to the current $J_\mu$ as conventional photons, 
but also have additional Lorentz-violating background interactions 
like those appearing in the Standard-Model Extension.
By studying these additional interactions,
signatures can be searched for that might ultimately
distinguish between a photon theory based on local Lorentz breaking
from that of conventional Einstein-Maxwell theory.

\section{Higgs Mechanisms}

Since there are two sets of broken symmetries (Lorentz and diffeomorphisms) 
there are potentially two associated  Higgs mechanisms. 
However,
in addition to the usual Higgs mechanism
(in which a gauge-covariant-derivative term generates a mass term
in the Lagrangian),
it was shown\cite{ks}
that an alternative Higgs mechanism can occur due to the
gravitational couplings that appear in the potential $V$.

First, it was shown 
that the usual Higgs mechanism involving the metric does not occur
\cite{ks}.
This is because the mass term that is generated by covariant derivatives
involves the connection,  which consists of derivatives of the metric
and not the metric itself.
As a result,
no  mass term for the metric is generated following the usual Higgs prescription.
However, it was also shown that because of the form of the potential, 
e.g., $V=V(B_\mu g^{\mu\nu} B_\nu + b^2)$, 
quadratic terms for the metric can arise, 
resulting in an alternative form of the Higgs mechanism
\cite{ks}.
These can lead to mass terms that can potentially modify gravity
in a way that avoids the van Dam, Veltmann, and Zakharov discontinuity.

In contrast, for the case of Lorentz symmetry, 
it is found that a conventional Higgs mechanism can occur
\cite{rbak05}.
In this case the relevant gauge field (for the Lorentz symmetry) 
is the spin connection.  
This field appears directly in covariant derivatives acting on local tensor
components,
and for the case where the local tensors acquire a vev,
quadratic mass terms for the spin connection can be generated
following a similar prescription as in the usual Higgs mechanism.
For example. in the bumblebee model, 
using a unitary gauge, 
the kinetic terms involving $B_{\mu\nu}$ generate quadratic mass terms
for the spin connection. 
However, a viable Higgs mechanism involving the spin connection can 
only occur if the spin connection is a dynamical field.  
This then requires that there is nonzero torsion and 
that the geometry is Riemann-Cartan.

\section{Summary and Conclusions}

In theories with spontaneous Lorentz violation, 
up to ten NG modes can appear. 
They can all be incorporated naturally in the vierbein.  
For the vector bumblebee model, 
the Lorentz NG modes propagate like photons in an axial gauge. 
In principle, two Higgs mechanisms can occur,
one associated with broken diffeomorphisms,
the other with Lorentz symmetry.
While a usual Higgs mechanism (for diffeomorphisms)
does not occur involving the metric field,
an alternative Higgs mechanism can lead to the appearance of quadratic
metric terms in the Lagrangian.
If in addition the geometry is Riemann-Cartan, 
then a Higgs mechanism (for the Lorentz symmetry)
can occur in which the spin connection acquires a mass.

\section*{Acknowledgments}
This research has been supported by
NSF grant PHY-0554663.

\vfill

\end{document}